# The Trouble with "Puddle Thinking": A User's Guide to the Anthropic Principle

**Geraint F. Lewis[1*] and Luke A. Barnes[2]**

[1] Sydney Institute for Astronomy, School of Physics, A28,
The University of Sydney, NSW 2006, Australia
[2] Western Sydney University, Locked Bag 1797,
Penrith South, NSW 1797, Australia

\* Corresponding author.
E-mail: geraint.lewis@sydney.edu.au

## Abstract

Are some cosmologists trying to return human beings to the centre of the cosmos? In the view of some critics, the so-called "anthropic principle" is a desperate attempt to salvage a scrap of dignity for our species after a few centuries of demotion at the hands of science. It is all things archaic and backwards – teleology, theology, religion, anthropocentrism – trying to sneak back in scientific camouflage. We argue that this is a mistake. The anthropic principle is not mere human arrogance, nor is it religion in disguise. It is a necessary part of the science of the universe.

## Introduction

In the 1930s, the Nobel-Prize-winning physicist Paul Dirac was pondering strange coincidences between the fundamental numbers of nature (Dirac 1938). He worked out the ratio of the *electromagnetic* force to the *gravitational* force between an electron and proton in an atom and got a huge number: $10^{40}$. He also worked out the ratio of the *age of the universe* to the time it takes for *electrons to orbit* in an atom and got another huge number: $10^{39}$. Curiously, these numbers are similar. Maybe it's just a coincidence, or maybe — Dirac thought — it's a clue to deeper laws of nature.

In the early 1960s, astronomer Robert Dicke compellingly argued that it was *neither* (Dicke 1961). He realised that there is something usual about Dirac's relation, something hiding inside one of the quantities: *us*. Like all of us, the universe is getting older. So, the *age of the universe* in Dirac's second ratio isn't a fundamental constant. It's the time between the beginning of the universe and *us, here, now, today*. Any account of the coincidence must consider how the Universe makes beings that are capable of measuring its age.

Dicke realised that we cannot be living at any random time in the universe. Firstly, in its youth, the cosmos was a featureless sea of the simplest atoms, hydrogen and helium. The elements needed for life — from the carbon that provides the backbone for organic molecules, to the calcium that provides the backbone for our backbones — are formed in nuclear reactions at the hearts of stars and are recycled by stellar winds and supernova explosions into planets, and ultimately life. Secondly, in the dim and distant future, most of the stars have died, and the energy to sustain life becomes rare. The building blocks for planets and people are entombed in the ever-cooling cores of stars or inside black holes. Life, in this distant future universe, would be precarious, and probably much rarer than today.

Putting these two facts together, *given that life exists at all*, we should not be surprised to find that when we measure the age of the universe, we get an answer that is greater than (but not too much greater than) the lifetime of a star. When we express this relation in terms of the fundamental constants (using a simple model for stars), we get Dirac's coincidence.

It is a mistake to think that Dicke is saying that our time in the universe is "special," that "our Universe stands at a 'golden interval', neither too young nor too old, but just right."[1] Rather, Dicke is employing a basic principle of the scientific method: what you observe depends on what you are looking *at* **and** what you are looking *with*. When it comes to the universe, we are not Dr Frankenstein, setting up our scientific equipment when and where we please. We are the monster: we have woken up in the middle of the contraption that made us and are trying to understand how it all works.

## Looking through our eyes

The natural question for cosmologists and physicists to ask next is: what else about our universe could be explained in this way? What combination of fundamental laws and our necessarily limited perspective best accounts for our observations of the universe?

In search of the answer, physicists delved into the deepest properties of nature, including the masses of the fundamental particles and the strengths of the fundamental forces. By considering other hypothetical universes, it was found that slight deviations in these fundamental properties often result in dead and sterile universes that lack the complexity necessary for life (for a recent review, see Adams 2019). This is known as the cosmological fine-tuning problem: the ability of the fundamental laws of our universe to provide the right conditions for life of any conceivable kind is a seemingly very rare talent indeed. As summarised in our recent book *A Fortunate Universe: Life in a finely tuned cosmos* (2016), many small changes have disastrous effects. If the strong force were slightly weaker or the fundamental masses slightly heavier, the periodic table would not exist. If gravity were weaker or the universe expanded too fast, matter would not form into stars to forge elements, or indeed make any structure at all. Such a universe would be too simple, too short-lived, or too empty to ever host life.

Note well: we have arrived here without any assumptions about human specialness or religious jiggery-pokery. Saying that the universe is "fine-tuned for life" is not to say that it has a fine-tuner! It is only to say that there is something rare about the physical parameters that life requires. We're just doing science. Fine-tuning for life has been studied by physicists for decades, using the best theoretical tools available, and published in peer-reviewed journals.

## Other Life-forms and Other Universes

Wait a minute, we hear you say. How can you make such sweeping statements about life and universes when we don't have a good definition of what life is, and we don't know what other universes are even possible?

---

[1] "Anthropic arrogance," David P. Barash, Aeon: aeon.co/essays/why-a-human-centred-universe-is-not-a-humane-one

For the first objection, we reply that the fine-tuning for life is really the fine-tuning for the complexity required by life. We don't assume that another possible way the universe could have been is life-prohibiting because we couldn't live there. The kind of life-prohibiting disasters that await in other universes are the non-existence of chemistry, or indeed, any way at all to stick two particles together. Or a universe that ends before anything could stick together. Or a universe that expands so fast that no two things have any chance of sticking together. This is a long way from the debate over whether a virus is alive.

But how do we know that these other universes are possible? As the ANU's Charley Lineweaver has pointed out to us, "There is no fine-tuning if there are no knobs." But think about that claim for a moment. These other, life-prohibiting universes are perfectly mathematically consistent. So who took the knobs away? A deeper physical law? Great! What is it? And why is it a physical law that allows life forms, rather than one that doesn't? In the words of Carr and Rees (1979), "even if all apparently anthropic coincidences could be explained [by some presently unformulated physical theory], it would still be remarkable that the relationships dictated by physical theories happened also to be those propitious for life."

Perhaps something deeper than the laws of nature took the knobs away, like a metaphysical principle? Great! What is it? And why is it a metaphysical principle that allows life forms, rather than one that doesn't? And what a stunning comeback for armchair philosophy! Scientists have been toiling for centuries, learning about the universe by actually measuring it. But all this time, we could have been deriving the mass of the electron from some *a priori* philosophical principle with a deep affinity for the number $4.185463 \times 10^{-23}$ (the electron mass in Planck units).

## Whence the Anthropic Principle?

The term "anthropic principle" comes from a presentation by astrophysicist Brandon Carter in 1973, at a celebration of Copernicus's 500th birthday. Building upon the insights of Dicke and others, Carter argued that our position in time and space must be taken into account in our scientific theorising about the world, noting that:

> Although our situation is not necessarily *central*, it is inevitably privileged to some extent.

Carter is echoing Dicke's insight: there are times and places in our universe where life is overwhelmingly more likely to exist, and so our perspective on the universe is *necessarily* limited. This is what Carter called the *weak anthropic principle*.

Carter also proposed a *strong anthropic principle*:

> The Universe (and hence the fundamental parameters on which it depends) must be as to admit the creation of observers within it at some stage.

This principle is liable to be misunderstood due to the word "must." Its sense here is consequential, as in "there is frost on the ground, so it must be cold outside." We are physical life forms capable of measuring the universe, but not all fundamental laws allow for such things. Carter's strong anthropic principle is *not* proposing that our existence *causes* the universe's fundamental properties, or that any deep metaphysical principle or divine being was involved.

Here's where the confusion starts: others have not followed Carter. In 1986, physicists John Barrow and Frank Tipler published the influential book, *The Cosmological Anthropic Principle.* They brilliantly explained how the overall properties of the cosmos, the details of the fundamental particles, and the forces that bind them together combine to produce the complexity and energy necessary for life.

But on the anthropic principle, Barrow and Tipler muddied the waters by giving the *same* term a *different* definition. They proposed a weak anthropic principle that combines Carter's strong and weak principle:

> The observed values of all physical and cosmological quantities are not equally probable but they take on values restricted by the requirement that there exist sites where carbon-based life can evolve and by the requirement that the Universe be old enough for it to have already done so.

This is Dicke's insight. So far, so good. But Barrow and Tipler proposed a new strong anthropic principle, one that *sounds* similar to Carter's strong principle, except that the word "must" is now given full speculative licence. Perhaps, they say, the universe has a designer, or "observers are necessary to bring the Universe into being." This version of the strong anthropic principle is metaphysical.

So, now we have two versions of the weak anthropic principle and two versions of the strong anthropic principle, those from Carter and those from Barrow and Tipler. Confusion was inevitable.

In addition, Barrow and Tipler added yet more "anthropic principles," such as the Final Anthropic Principle: "Intelligent information-processing must come into existence in the universe, and, once it comes into existence, it will never die out." The dominant form of life over the history of the universe would be some kind of über-computer, digital consciousnesses enjoying an everlasting virtual reality paradise. This is, to put it mildly, speculative.

Thus, Carter's important and necessary idea has become both associated with disreputable and speculative company. Understandably, many scowl whenever the anthropic principle is mentioned. The feeling is that the anthropic principle is at best tautological and invoking it to explain any feature of our Universe is "the last refuge of the scoundrel." And, at worst, the principle is untestable conjecture.

## Are we a puddle in a hole?

The question is: what do we do with the fine-tuning of the universe for life? Does (Carter's) anthropic principle explain why a life-permitting universe exists?

Douglas Adams, in his posthumously published book *The Salmon of Doubt* (2002), famously lampooned the attempt to argue from the features of our environment to any greater cosmic purpose:

> This is rather as if you imagine a puddle waking up one morning and thinking, "This is an interesting world I find myself in — an interesting hole I find myself in — fits me rather neatly, doesn't it? In fact it fits me staggeringly well, must have been made to have me in it!" This is such a powerful idea that as the sun rises in the sky and the air heats up and as, gradually, the puddle gets smaller and smaller, frantically hanging on to the notion that everything's going to be alright, because this world was meant to have him in it, was built to have him in it; so the moment he disappears catches him rather by surprise. I think this may be something we need to be on the watch out for.

Adams is a favourite of ours, and on one level, this pithy story nicely illustrates the approach of Dicke: be mindful of the process that made you when you try to understand your environment. We are not detached observers of the universe, but are part of it, formed and shaped by its physical laws and our immediate cosmic habitat. *Given that we exist at all*, we should not be surprised that life-sustaining environments exist, even though the majority of the universe is inhospitable to human life.

But some have pushed the accusation of "puddle thinking" too far, supposing that it solves the fine-tuning puzzle.

Consider more closely the puddle's reasoning. Let's name our puddle *Doug*. He has noticed a precise match between two things: 1) his shape and 2) the shape of the hole in which he lives. Doug is amazed! What Doug doesn't know is that, given A) the fluidity of water, B) the solidity of the hole, and C) the constant downward force of gravity, he will *always* take the same shape as his hole. If the hole had been different, his shape would adjust to match it. Any hole will do for a puddle.

This is precisely where the analogy fails: any universe will *not* do for life. Life is not a fluid. It will not adjust to any old universe. There could have been a completely dead universe: perhaps one that lasts for 1 second before recollapsing or is so sparse that no two particles ever interact in the entire history of the universe.

Think about the real explanation to Doug's observation: A (fluid water) + B (solid hole) + C (gravity). If the puddle analogy applies to fine-tuning, what corresponds to A+B+C? What explains the match between what our universe does and what life requires? The puddle analogy doesn't say. Invoking the puddle against fine-tuning is essentially saying "perhaps a solution exists." Well, OK, sure, thanks for that, but what could that solution be? Maybe you could go one step further by filling in the blank in the following claim: a universe permits the possible existence of life because ____________.

Here's the thing: Doug is *right* to think that the match between his shape and the shape of the hole is worthy of explanation. He is not arrogant to look for an explanation. He would be unwise to dismiss *without good reason* the supposition that he is designed for the hole; after all, if Doug talked to his pals Lock and Key, they too would tell him of their remarkable matching shapes. We understand the puddle; we understand a lock and key; we want to understand fine-tuning for life. But "puddle thinking" is often used as an excuse to dismiss fine-tuning as unworthy of our attention at all. Even Doug knows better than that!

## A fine-tuning puzzle

The real conundrum of life in the Universe is not: given that we are here, why do we find ourselves in a universe with the conditions that allow us to be here? The puzzle is: why does a universe with the ability to support life exist at all?

This question is uncomfortable for many because it takes us to the edge of physics. If we ever uncover the ultimate properties of physical reality, we will have reached the end of physical explanations. Either the universe is as it is for no reason, or we must look for a reason beyond physics. The debate is unavoidable but necessarily philosophical.

Some invoke a divine mind, a "fine-tuner" who configured the universe to allow us to be here. Perhaps ours is a synthetic universe, whose conditions were chosen by a programmer who wants to simulate an interesting universe.

Another live option is the multiverse, the notion that our universe is one of many, each with their own physical laws and conditions. In many proposed models of a multiverse, most universes are dead and sterile, but with enough spins of the cosmic roulette wheel, the right conditions for life should show up somewhere. We should not be surprised to find ourselves in one with physical conditions that allow us to be here.

At the moment, the multiverse is a rough sketch of a scientific theory, or more exactly, a collection of sketches. If we had a rigorous multiverse theory, we could predict the variety of generated universes and see whether our universe is rare or common. Just as importantly, we could ask: are life-permitting universe generators as fine-tuned as life-permitting universes?

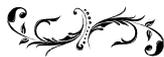